\documentclass{acm_proc_article-sp}
\usepackage{amsfonts}
\usepackage{graphicx}

\newtheorem{prop}{Proposition}
\newtheorem{lem}{Lemma}
\newtheorem{rem}{Remark}

\begin{document}
\title{Online Learning of Aggregate Knowledge about Non-linear Preferences Applied to Negotiating Prices and Bundles}
\numberofauthors{3}

\author{
\alignauthor D.J.A.~Somefun\\
    \affaddr{Center for Mathematics and Computer Science (CWI)}\\
    \affaddr{P.O. Box 94079, 1090 GB Amsterdam,The Netherlands}\\
    \email{koye@cwi.nl}
\alignauthor T.B.~Klos\\
    \affaddr{Center for Mathematics and Computer Science (CWI)}\\
    \affaddr{P.O. Box 94079, 1090 GB Amsterdam,The Netherlands}\\
    \email{tomas@cwi.nl}
\alignauthor J.A.~La~Poutr\'e\\
    \affaddr{Center for Mathematics and Computer Science (CWI)}\\
    \affaddr{and Technical Univ. Eindhoven}\\
    \affaddr{School of Techn. Management}\\
    \affaddr{P.O. Box 513, 5600 MB Eindhoven, The Netherlands}\\
    \email{hlp@cwi.nl}
}


\date{}
\maketitle

\begin{abstract}
In this paper, we consider a form of multi-issue negotiation where a shop negotiates both the contents and the price of bundles of goods with his customers. We present some key insights about, as well as a procedure for, locating mutually beneficial alternatives to the bundle currently under negotiation. The essence of our approach lies in combining aggregate (anonymous) knowledge of customer preferences with current data about the ongoing negotiation process.

The developed procedure either works with already obtained aggregate knowledge or, in the absence of such knowledge, learns the relevant information online. We conduct computer experiments with simulated customers that have \emph{nonlinear} preferences. We show how, for various types of customers, with distinct negotiation heuristics, our procedure (with and without the necessary aggregate knowledge) increases the speed with which deals are reached, as well as the number and the Pareto efficiency of the deals reached compared to a benchmark.
\end{abstract}

\section{Introduction}\label{sec:intro}

Combining two or more items and selling them as one good, a practice called bundling, can be a very effective strategy for reducing the costs of producing, marketing, and selling products \cite{Baumol/Etal:1987}. In addition, and maybe more importantly, bundling can stimulate demand for (other) goods or services \cite{Schmalensee:1984,Bakos/Brynjolfsson:1999}. To stimulate demand by offering bundles of goods, requires knowledge of customer preferences. Traditionally, firms first acquire such aggregate knowledge about customer preferences, for example through market research or sales data, and then use this knowledge to determine which bundle-price combinations they should offer. Especially for online shops, an appealing alternative approach would be to \emph{negotiate} bundle-price combinations with customers: in that case, aggregate knowledge can be used to facilitate an interactive search for the desired bundle and price. Due to the inherently interactive characteristics of negotiation, such an approach can very effectively adapt the configuration of a bundle to the preferences of a customer. A high degree of bundle customization can increase customer satisfaction, which may lead to an increase in the demand for future goods or services.

In this paper, we present an approach that allows a shop to make use of aggregate knowledge about customer preferences. Our procedure uses aggregate knowledge about \emph{many} customers in bilateral negotiations of bundle-price combinations with \emph{individual} customers. Negotiation concerns the selection of a subset from a collection of goods or services, viz.\ the bundle, together with a price for that bundle. Thus, the bundle configuration---an array of bits, representing the presence or absence of each of the shop's goods and services in the bundle---together with a price for the bundle, form the negotiation issues. In theory, this is just an instance of multi-issue negotiation. Like the work of~\cite{Klein2003,faratinSierraJennings:2003,Ehtamo2001,somefunGerdingBohtePoutre:2003-amec}, our approach tries to benefit from the so-called win-win opportunities offered by multi-issue negotiation, by finding mutually beneficial alternative bundles during negotiations. The novelty of the approach, however, lies in the use of aggregate knowledge of customer preferences. We show that the bundle that represents the highest `gains from trade' Pareto-dominates all other bundles within a certain collection of bundles.\footnote{The gains from trade for a bundle are equal to the customer's `valuation' of the bundle minus the shop's valuation of the bundle, which is his (minimum) price (cf. \cite{Mas-Collel/Etal:1995}).}$^,$\footnote{An offer constitutes a Pareto improvement over another offer whenever it makes one bargainer better off without making the other worse off. A bundle $b'$ `Pareto-dominates' another bundle $b$ whenever switching from $b$ to $b'$ results in a Pareto improvement (cf. \cite{Mas-Collel/Etal:1995}).}\newcounter{paretoFootnote}\setcounter{paretoFootnote}{\thefootnote} Based on this important insight, we develop a procedure for combining aggregate knowledge of customer preferences with data about an ongoing negotiation process with 1 customer, to find alternative bundles that are likely to lead to high Pareto improvements. Note that due to the use of aggregate data only, our approach does not necessitate infringement of customers' privacy. 

The procedure we developed requires a process in the \textbf{foreground} and one in the \textbf{background}. The foreground process uses aggregate knowledge about customer preferences to recommend promising alternative bundles during ongoing negotiations with customers. Intuitively, the idea for the process in the foreground is that, whenever the shop decides to stop bargaining about a bundle $b$ and to switch to an alternative bundle, he will choose from a `neighborhood' of $b$, the bundle that looks promising in the sense that it has the highest conditionally expected gains from trade. The background process obtains the necessary aggregate knowledge about customer preferences. Based on this knowledge it estimates for each bundle the expected gains from trade, conditional on what the ongoing negotiation process reveals about the current customer. 

With respect to the background process we consider two cases. In the first case, we do not explicitly consider the background process: the shop already possesses the necessary aggregate knowledge. The shop may have obtained this aggregate knowledge by having access to expert knowledge or by collecting historical sales data and mining this data \emph{offline}. The main purpose of this case is to highlight the value of the foreground process given sufficient aggregate knowledge, and to provide an upper bound for the second case. In the second case, we explicitly consider the background process: the shop does not have any a priori knowledge of customer preferences. Instead he learns about customer preferences \emph{online} by interpreting individual customers' responses to the shop's proposals for negotiating about alternative bundles. This allows the shop to make progressively better estimations of the expected gains from trade.

To ensure that bundling can stimulate demand for (other) goods or services we conduct computer experiments with simulated customers that have \emph{nonlinear} preferences: i.e., a customer's valuation for a bundle of goods may be higher (or lower) than the sum of the customer's valuations for the individual goods. In our experiments, we consider the foreground process both with and without the aggregate knowledge already being available. In the absence of aggregate knowledge, the background process will learn the relevant information online. We show how, for various types of customers---with distinct negotiation heuristics---the foreground process (both with and without the necessary aggregate knowledge) increases the speed with which deals are reached, as well as the number and the Pareto efficiency of the deals reached compared to a benchmark. Moreover, through time, the performance of the foreground process without a priori information approaches the procedure that already possesses the necessary aggregate knowledge.

The subproblem of just finding a good (or better) bundle configuration can be seen as a form of recommending \cite{resnickVarian:1997-cacm}, if we do not consider the negotiation and pricing aspects. The general subject of bundling has received a lot of attention recently, especially in the context of online information goods~\cite{kephartFay:2000-acmec,Somefun/Poutre:2003,altinkemerJaisingh:2002-ieeewecwis,Bakos/Brynjolfsson:1999,chuangSirbu:1999-iep}. The issue of finding the appropriate bundles is, however, not limited to information goods. It also occurs outside of the realm of information goods, where a number of aspects of a complex product can be selected, such as for a PC~\cite{costerGustavssonOlssonRudstrom:2002-rpec}, a trip~\cite{sooLiang:2001-cia}, or photography equipment~\cite{molina:2001-wi}. Until now, this has not been considered as part of a negotiation process, to the best of our knowledge.

For numerous real word applications---like the above examples of selecting aspects of a complex product---the number of individual goods to be bundled, $n$, is relatively small. In this paper we will also only consider small values of $n$ (say $n\leq 10$), for which aggregate knowledge still greatly facilitates the process of finding attractive alternative bundles during a negotiation process. For example, with $n=10$ there are $2^n -1=1023$ possible bundle configurations, so facilitating the search process among all those bundles is highly valuable. 

This paper builds on and significantly extends the idea, developed in a preceding paper \cite{Somefun/Etal:2004a}, to negotiate over bundles and prices using aggregate knowledge. The scope of the earlier paper is limited to the foreground process; the necessary aggregate knowledge is assumed to be already available. This approach is warranted because the paper focuses on additively separable preferences (i.e., a customer's valuation for a bundle is always equal to the sum of her valuations for the individual goods comprising the bundle). With additive separability it suffices to learn the conditional expected gains from trade for the individual goods (cf. \cite{Somefun/Etal:2004a}), which greatly simplifies the problem of learning the required aggregate knowledge. In this paper we consider non-linear customer preferences, for which learning the desired aggregate knowledge can be very difficult. For example, it may be very difficult to determine the conditionally expected gains from trade by collecting historical sales data and mining those data offline. It requires that the sales data reveals the correlation between customers' valuations for the various bundles. Such high quality data may not be readily available, especially when at the same time customers' privacy should be respected, as we assume in this paper. By interpreting customers' online responses to the shop's proposals for negotiating about alternative bundles, our background process circumvents these difficulties.

The next section provides a high-level overview of the interaction model. In Section~$\ref{sec:usermodel}$ we introduce relatively mild conditions on the customers' and the shop's preferences. Based on these conditions, Section~$\ref{sec:model}$ develops a procedure for finding the most promising alternative bundles. In order to test the performance of our system, we used it in interactions with simulated customers. In Section~$\ref{sec:determiningGains}$ we discuss how the necessary aggregate knowledge of customer preferences is learned online. Section~$\ref{sec:simulation}$ presents our computer experiments and discusses the results. Conclusions follow in Section~$\ref{sec:discussion}$.

\section{Overview}\label{sec:overview}

This section gives an overview of the interaction between the shop and the customer, as they try to negotiate an agreement about the price and the composition of a bundle of goods. The shop sells a total of $n$ goods, each of which may be either absent or present in a bundle, so that there are $2^n - 1$ distinct bundle-configurations containing at least $1$ good. In the current paper, we use $n = 10$. A negotiation concerns a bundle (configuration), together with a price for that bundle, and it is conducted in an alternating exchange of offers and counter offers \cite{Rubinstein:1982}, typically initiated by the customer. An example of such a practice may involve the sales of bundles of news items in categories like politics, finance, economy, sports, arts, etc.

We develop a procedure that a shop can use to find mutually beneficial alternative bundles during the negotiation about a given bundle, so that alternative bundles may be recommended whenever the negotiation about the given bundle stalls. Specifically, the procedure finds \emph{Pareto improvements} by changing the bundle content.\footnotemark[\theparetoFootnote] It uses information specific to the current negotiation process as well as aggregate knowledge (obtained from the analysis of sales data, (anonymous) data on previous and current negotiations, market research, or expert knowledge). The ongoing negotiation is analyzed to determine \emph{when} an alternative bundle is needed, and both the ongoing negotiation process and the aggregate knowledge are used to assess \emph{which} bundle to recommend.

A customer can \emph{explicitly} reject a suggested bundle by specifying a counter offer with a different bundle content (e.g., the previous one), and she can \emph{implicitly} reject a suggested bundle by offering a low price for it. In the current paper, only implicit rejection is allowed: customers only specify the bundle content for the opening offer, and thereafter only the shop can change the bundle content of an offer. This is to ease the description of our model and solutions. The possibility for customers to explicitly reject or change the bundle content can be easily incorporated in our model and solutions, however.

\begin{figure*}[ht]
\center
\centering \resizebox{.75\textwidth}{!}{\includegraphics*{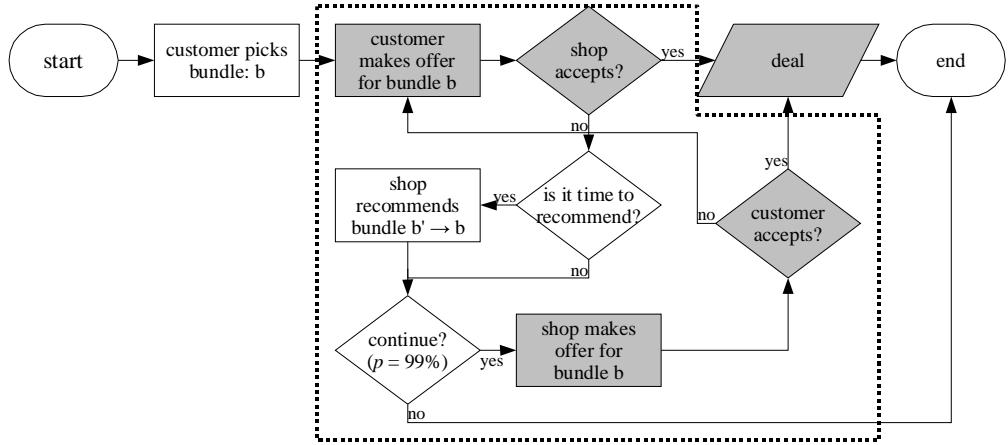}}
\caption{A flowchart describing the integration of recommendation in a shop and a customer's alternating exchange of offers and counter offers.}
\label{fig:flowchart} 
\end{figure*}

Figure~$\ref{fig:flowchart}$ provides a high-level overview of the interaction between a shop and a customer. The shaded elements are part of the actual negotiation---the exchange of offers.
The process starts with the customer indicating her interests, by specifying the bundle they will initially negotiate about. After that, they enter into a loop (indicated by the dotted line) which ends only when a deal is made, or with a $2\%$ exogenous probability. (We do not model bargainers' impatience explicitly; therefore we need an exogenous stopping condition, which specifies the chance of bargaining breakdown.) In the loop, the customer makes an offer for the current bundle $b$, indicating the price she wants to pay for it. The shop responds to this offer either by accepting it, or by considering a recommendation. In any case, conditional upon the $98\%$ continuation probability, the shop also makes an offer, either for the current bundle $b$ or for a recommended bundle $b'$ (which then becomes the current bundle). 

In the model, the \emph{valuations} of the customers and the shop are expressed as \emph{monetary values}. The \emph{utilities} of deals are expressed as strictly monotonic one-dimensional transformations of valuations. In the simplest form, this would be the difference between the valuation of the bundle and the negotiated price. The agents are interested in obtaining a deal with optimal utility (``net monetary value''). See Section~$\ref{sec:usermodel}$ for details.

\section{Preference Model}\label{sec:usermodel}

\subsection{Informal Discussion}

The essence of our model of valuations and preferences lies in the assumption that customers and the shop order bundles based on their \emph{net monetary value}; the bundle with the highest net monetary value is the most preferred bundle. A customer's net monetary value of a bundle is equal to the customer's valuation of the bundle (expressed in money) minus the bundle price and the shop's net monetary value is equal to the bundle price minus the shop's bundle valuation (also expressed in money).

Given the above assumption and the assumption that a customer wants to buy at most one bundle (within a given time period), Section~$\ref{subsection:formal_discussion}$ shows that any deal involving the bundle with the highest \emph{gains from trade} is Pareto efficient. We can now specify which is the best bundle for the shop to advise: faced with the problem of recommending one bundle out of a collection of bundles, the ``best'' bundle to recommend is the bundle with the highest expected gains from trade; this bundle Pareto dominates all other bundles. (Section~$\ref{subsection:formal_discussion}$ can be skipped upon first reading.)

\subsection{Formal Discussion}\label{subsection:formal_discussion}

Before being able to more formally state the results, some notation is necessary. Let $N\subset \mathbb{N}$, with $n=|N|$, denote the collection of $n$ individual goods and $2^N$ denote the power set of $N$ (i.e., the collection of all subsets of $n$), then $B=2^N\setminus\{\emptyset\}$ denotes the collection of all possible bundles. Furthermore, let $P = \mathbb{R}$ denote the collection of all possible bundle prices.\footnote{Negative prices may not be realistic, but we want to make as few behavioral assumptions as possible. For the results the possibility of negative prices is not problematic (see Footnote~$\ref{footnote:pos_prices}$).} The customer and the shop attach the monetary values of $v_c(b)$ and $v_s(b)$, respectively, to a bundle $b\in B$ (with $v_c(b), v_s(b) \in P$). The function $x_j : B \times P \mapsto \mathbb{R}$ with $j \in \{c,s\}$ denotes the \emph{net monetary value} for bundle $b$ at price $p$: $x_c(b, p) = v_c(b) - p$ and $x_s(b,p)= p - v_s(b)$ denote the customer's and the shop's net monetary values, respectively.

We assume that the customer's and the shop's utility for consuming bundle $b$ at price $p$, denoted by $u_j(b, p)$ with $j\in \{c,s\}$, can be expressed as the composition function $g_j \circ x_j(b, p)$ with $g_j:\mathbb{R} \mapsto \mathbb{R}$. For $g_j$ we assume that $\frac{dg_j(x)}{dx} > 0 \mbox{ for all }x\in \mathbb{R}$. Thus we have that $u_j(b,p)=g_j(x_j(b,p))$ and since $g_j$ is a strictly increasing function, we can assume without loss of generality that $u_j(b,p)=x_j(b,p)$ (cf. \cite{Mas-Collel/Etal:1995}).  

Given the customer's and shop's monetary values, we define a useful subset $B^*$ of $B$ as follows: $B^* \equiv \arg\max_{b\in B}( v_c(b) - v_s(b))$, that is, $ B^*$ represents the collection of bundles with the highest possible gains from trade (across all possible bundles). We are now ready to introduce the following proposition.

\begin{prop}\label{prop:paretoEffBundle}
A deal $(b, p)$ with $b \in B$ and $p \in P$ is Pareto efficient if and only if $b\in B^*$.
\end{prop}

\begin{rem}
A deal $(b, p)$ is Pareto efficient if there is no $(b', p')$ such that $u_j(b, p) \leq u_j(b',p')$ for all $j\in \{c,s\}$ and the inequality is strict for at least one $j$.
\end{rem}

Proposition~$\ref{prop:paretoEffBundle}$ means that a deal is Pareto efficient if and only if it entails a bundle with the highest possible gains from trade. For the proof of this proposition the following lemma is very useful.
\begin{lem}\label{lemma:inequality}
For any two deals $(b^*, p^*)$ and $(b, p)$ with $p^*, p \in P$, $b^* \in B^*$, and $b \in B \setminus B^*$ we have $x_c(b, p) < x_c(b^*, p^*)$ or $x_s(b, p) < x_s(b^*, p^*)$.  
\end{lem}

\begin{proof}
We prove the above lemma by contradiction. Suppose that for any $b^* \in B^*$ and $b \in B \setminus B^*$ we have $x_c(b, p) \geq x_c(b^*, p^*)$ and $x_s(b, p) \geq x_s(b^*, p^*)$. A necessary condition for this to hold is that $v_c(b) - v_s(b) \geq v_c(b^*) - v_s(b^*)$. However, $b^* \in B^*$ and $b \in B \setminus B^*$ means, by definition of $B^*$, that $v_c(b) - v_s(b) < v_c(b^*) - v_s(b^*)$.
\end{proof}

We are now ready to prove Proposition~$\ref{prop:paretoEffBundle}$.

\begin{proof}

\begin{enumerate}
\item (If) Pick any $j \in \{c, s\}$. Suppose that $j$'s position improves by moving from any deal $(b, p)$ with $b \in B^*$ to $(b', p')$, that is, $u_j(b, p) < u_j(b', p')$. It then suffices to show that the opponent denoted by $j'$ will always be made worse off, that is, $u_{j'}(b, p) > u_{j'}(b', p')$. From the properties of $g_j$ and $g_{j'}$ it follows that a bargainer's position improves/worsens whenever the net monetary value increases/decreases. Since $j$'s position improves, it follows from Lemma~$\ref{lemma:inequality}$ that $j'$ is made worse off whenever $b \in B \setminus B^*$. Moreover, if $b,b' \in B^*$ then the gains from trade remain unchanged, hence $j'$ is made worse off.
\item (Only if) We will prove this part by contradiction. Suppose that $b \notin B^*$ with the price being any $p \in P$. Pick any $b' \in B^*$ and set the bundle price to $p' = p + v_s(b') - v_s(b)$, so that $p' - v_s(b') = p - v_s(b)$. It follows from $p \in P$ that $p' \in P$ (recall that $P = \mathbb{R}$)%
\footnote{\label{footnote:pos_prices}If we choose to a priori rule out $p < 0$ and $v_j(b) < 0$ (for $j \in \{c, s\}$ and all $b \in B$), then $p \geq v_s(b)$ should hold because otherwise the shop will not be willing to sell the bundle in the first place. Consequently, $p' \in P$ still holds.}
and the properties of $g_s$ that the shop is indifferent between the deals $(b, p)$ and $(b', p')$. Also, it follows from Lemma~$\ref{lemma:inequality}$ and the properties of $g_{c}$ that the customer is made better off. That is, any $b' \in B^*$ Pareto dominates $b \notin B^*$. Thus $b \notin B^*$ cannot be a Pareto efficient solution.   
\end{enumerate}

\end{proof}

\section{The Foreground Process}\label{sec:model}

The idea is to develop a mechanism for a shop to find Pareto improvements by changing the bundle content during a negotiation. The mechanism we propose contains two subprocedures. The first procedure monitors the negotiation process and tells the shop \emph{when} to recommend, at which time the second procedure tells the shop \emph{what} to recommend, by generating recommendations based on aggregate knowledge and the ongoing negotiation process. Figure~$\ref{fig:flowchart}$ shows the interaction between these two procedures; they are discussed in more detail in Sections~$\ref{subsec:model-when2advise}$ and $\ref{subsec:model-recommending}$, respectively.

\subsection{Deciding \emph{When} to Recommend}\label{subsec:model-when2advise}

The shop needs a procedure for deciding when he should recommend negotiating about a different bundle. The obvious input for this decision is the progress of the current negotiation process, which can be described as a sequence of offers and counteroffers. An offer $O$ contains a bundle definition and a price: $O = (b, p)$ with $b \in B$ and $p \in P$. ($B$ and $P$ denote the collections of all possible bundles and prices, respectively.) Let $h = (O(1), O(2), \ldots, O(k))$ denote a finite history of offers ($k$ is a natural number), where $O(i + 1)$ is the counter offer for $O(i)$, for all $i < k$. Furthermore, let $H$ denote the universe of all possible finite offer sequences (thus $h\in H$). The problem of when to advise can now be specified as the mapping $f: H \mapsto \{\mathrm{now}, \mbox{not now}\}$, where ``$\mbox{(not) now}$'' means: (don't) recommend a new bundle now.

We construct a heuristic for $f$ based on the assumption that there is a probability of not reaching a deal with a customer (e.g., a break off, endless repetition, or deadline): the longer the negotiation is expected to take, the less likely a deal is expected to become. Furthermore, as a deal becomes less likely, the incentive for the shop to recommend negotiation about an alternative deal should increase. Given the shop's bargaining strategy, our heuristic then extrapolates the time the current negotiation process will need to reach a deal, from the pace with which the customer is currently giving in. More precisely, if we let $O = (b, p)$ and $O' = (b, p')$ denote the customer's current and previous offers for bundle $b$, then $\Delta t$, the predicted remaining number of negotiation rounds necessary to reach a deal, is defined as follows:
\begin{equation}
\Delta t= \frac{v_s(b) - p}{p - p'},
\end{equation} 
where $v_s(b)$ denotes the shop's monetary value for bundle $b$. The higher $\Delta t$, the higher the likelihood of a recommendation. Specifically, the probability of a recommendation depends on $\Delta t$ as follows:
\[
pr_{\mathrm{recommendation}} = 1 - \exp(-0.25 \Delta t),
\]
which means that the probability that the shop recommends an alternative bundle approaches 1 as $\Delta t$ increases.

\subsection{Deciding \emph{What} to Recommend}\label{subsec:model-recommending}

Our mechanism combines aggregate knowledge (obtained from the analysis of sales data, (anonymous) data on previous and current negotiations, market research, or expert knowledge) with data about the ongoing bargain process, to recommend bundles to customers while negotiating with them. Suppose, for example, that a customer offers to buy a bundle $b$ at a price $p$. Whenever a recommendation is needed (see Section~$\ref{subsec:model-when2advise}$) the idea is to select from within the ``neighborhood'' of bundle $b$, the bundle $b'$ that maximizes $E[v_c(b') - v_s(b') | v_c(b) \geq p]$: the expected gains from trade for bundle $b'$, given that the customer is willing to pay at least $p$ for bundle $b$. (To simplify notation we will write $E[ \cdot | b]$ instead of $E[ \cdot | v_c(b) \geq p]$.) Since the shop knows its own monetary value for bundle $b'$, $v_s(b')$, the aim is really to maximize $E[v_c(b') | b] - v_s(b')$. The difficulty here lies in estimating the customer's expected valuation of bundle $b'$:
\begin{equation}
E[v_c(b') | b] = \sum_{i \in P} i\cdot pr(v_c(b')=i| b),  
\end{equation}
where $pr(v_c(b') = i | b)$ denotes the probability that the customer's valuation for bundle $b'$ is equal to $i$, given that she is willing to pay at least $p$ for bundle $b$. In Section~$\ref{sec:determiningGains}$ we propose an online learning mechanism for determining this estimation (the background process mentioned in Section~$\ref{sec:intro}$. It is, however, instructive to first discuss the recommendation mechanism in some more detail (i.e., assuming that the expectations are already known).


A customer initiates the negotiation process by proposing an initial bundle and offering an opening price: let $O(0)=(b, p)$ denote the customer's opening offer (with $b \in B$ and $p \in P$). The shop stores the bundle proposed by the customer as (his assessment or estimation of) the customer's ``interest bundle,'' in the neighborhood of which the shop searches for promising alternative bundles to recommend if, at any time, the shop decides he should make a recommendation (see Section~$\ref{subsec:model-when2advise}$). This neighborhood of bundle $b$, $\mathit{Ng}(b)$, is defined as follows.
\begin{eqnarray}\label{b_advice_set}
\nonumber 
  \mathit{Ng}(b) & \equiv & \{b' \in B : b' \subset b\mbox{ and } |b'| + 1 =|b|\\
   & &\mbox{or }  b' \supset b\mbox{ and } |b'| - 1 =|b| \},
\end{eqnarray}
In other words, $\mathit{Ng}(b)$ contains the bundles which, in binary representation, have a Hamming distance to $b$ of $1$.\footnote{Remember that each bundle can be represented as a string containing $n$ bits indicating the presence or absence in the bundle, of each of the shop's $n$ goods.} The advantage of advising bundles within the neighborhood of $b$ is that the advice is less likely to appear haphazard.

Having defined a bundle's neighborhood, let the ordered set $A$ denote the so-called ``recommendation set,'' obtained by ordering the neighborhood $\mathit{Ng}(b)$ on the basis of the estimated expected gains from trade of all the bundles $b'$ in bundle $b$'s neighborhood, $\hat{E}[v_c(b') | b] - v_s(b')$, where $\hat{E}$ denotes the estimation of $E$. 

To recommend a bundle $b_k$ (the $k^\mathit{th}$ recommendation, with $k \geq 1$), our mechanism removes the first bundle from $A$, adds a price to it and proposes it as part of the shop's next offer. Depending on the customer's counter offer for bundle $b_k$, the current advice set may be replaced: if the customer's response is very promising (to be defined below) $A$ will be emptied, bundle $b_k$ will be taken as the customer's \emph{new} interest bundle (in the neighborhood of which the search continues), and the bundles in the neighborhood of $b_k$ are added to $A$.

To specify this in more detail, let $O^c_t$ denote the sequence of offers placed by the customer up until time $t$, and let $\mathit{max}(O^c_t)$ specify the customer's past offer with the highest net monetary value from the shop's perspective. Then the shop will determine the impact of the $k^\mathit{th}$ recommendation by comparing the net monetary value of the customer's current offer $O(t + 1)$ with that of offer $\mathit{max}(O^c_t)$. For this purpose, the shop uses the function $\mathit{sign_{b,b'}}: \mathbb{R} \times \mathbb{R} \mapsto \{0, 1\}$. If we let $\mathit{max}(O^c_t) = (b, p)$ and the customer's current offer $O(t + 1) = (b', p')$, then
\begin{equation}
\mathit{sign}_{b, b'}(p, p') =
    \left\{
        \begin{array}{lll}
            1 & \mbox{if } p' - v_s(b') > p - v_s(b)\\
            0 & \mbox{otherwise}
        \end{array}
    \right..
\end{equation}
If $\mathit{sign}_{b, b'} (p, p') = 1$, then the shop's assessment of the customer's interest bundle is updated to be $b_k$: the customer's response is promising enough to divert the search towards the neighborhood of $b_k$. That is, the first element of $A$ becomes the bundle $b' \in \mathit{Ng}(b_k)$ with the maximum difference $\hat{E}[v_c(b') | b_k] - v_s(b')$, the second element of $A$ becomes the bundle $b''$ with the second highest difference $\hat{E}[v_c(b'') | b_k] - v_s(b'')$, and so on. 

In case $\mathit{sign}_{b,b'}(p, p')=0$, the shop will make the next recommendation. Before the shop makes the next recommendation however, he checks if the negotiation is currently about the interest bundle. If this is not the case he will first make an offer containing the interest bundle. Whenever this offer is not accepted by the customer the shop will make the next recommendation in the following round. Consequently we have the property that a recommendation is always preceded by an offer containing the interest bundle. (We will see in Section~$\ref{sec:bargaining_data}$ that this is a very useful property.)

\section{The Background Process}\label{sec:determiningGains}

The ordering of all bundles in the neighborhood of an interest bundle $b$ constitutes a crucial aspect of the recommendation mechanism described in Section~$\ref{subsec:model-recommending}$. Ideally, given an interest bundle $b$, the first bundle in the ordering has the highest expected gains from trade, the second bundle in the ordering has the second highest expected gain from trade, and so on so forth. So, as explained in Section~$\ref{subsec:model-recommending}$, we are interested in knowing $E[v_c(b') | b] - v_s(b')$ for all bundles $b'$ within the neighborhood of bundle $b$, where the difficulty lies in estimating the customer's expected valuation $E[v_c(b')|b]$. Expert knowledge may provide the shop with these estimations, but unfortunately such knowledge is often not available. We will therefore introduce an effective approach for online learning to ``correctly'' order the bundles in the neighborhood of the interest bundle.

\subsection{Using Bargaining Data}\label{sec:bargaining_data}

To order the bundles, the shop uses data on the current and past bargaining processes. More precisely, suppose the shop advises $b'$ given an interest bundle $b$ with the most recent customer offer of $O=(b,p)$ and that the customer responds with the counter offer $O'=(b', p')$. The shop then feeds the triples $<b,b',p'-p>$ and $<b',b,p-p'>$ as new training examples to an online learning mechanism.

The recommendation mechanism described in Section~$\ref{subsec:model-recommending}$ ensures that the customer's offers $O=(b,p)$ and $O=(b',p)$ are placed directly after one another; thus, as long as the customer's strategic misrepresentation of the underlying bundle values do not jump around too much from one trading period to the next, the misrepresentation in $p$ and $p'$ will roughly cancel each other out. Consequently, $p-p'$ will be a good indication of the difference in a customer's valuations of bundles $b$ and $b'$, $v_c(b) - v_c(b')$. (Similarly, $p'-p$ will be good estimation of $v_c(b') - v_c(b)$.)

Based on these training examples the learning mechanism, when given $<b,b'>$ combined with the shop's valuations for the two bundles, $v_s(b)$ and $v_s(b')$, predicts $\Delta gt(b',b) \equiv E[p' - v_s(b')- (p - v_s(b))|b]$: the expected difference in gains from trade, resulting from changing from bargaining about bundle $b$ to bargaining about bundle $b'$ (given that a customer expressed an interest in bundle $b$, as assumed above). To sort bundles in the vicinity of an interest bundle $b$ according to their expected gains from trade, it suffices to sort the bundles according to $\Delta gt(b',b)$.

\subsection{Complexity Issues}

Knowledge of the correlation between the values of the various bundle pairs is essential for correctly learning to order all bundles in the vicinity of an ``interest" bundle $b$. Given that the shop sells $n$ individual goods, there are $2^n -1$ possible bundles containing at least 1 good. Learning the correlation between all bundle pairs requires---worst case---comparing an order of ${(2^n)}^2$ bundle pairs. Clearly, for particular instances of the problem the complexity may be reduced significantly. Take for instance the situation, where the customer's valuation for a bundle is always equal to the sum of her valuations for the individual goods comprising the bundle. In that case it suffices to compare $n$ individual goods with $2^n$ bundles, reducing the complexity---worst case---to an order of $n\cdot 2^n$. In this paper we focus on the more general case, where a customer's bundle valuation may not be equal to the sum of her valuations for the individual goods comprising the bundle.

For this more general customer preference setting, searching in the neighborhood of the interest bundle has two advantages: besides making an advice less likely to appear haphazard, it significantly reduces the number of bundle pairs that need to be considered. Recall that we defined the neighborhood of a bundle $b$ as consisting of all bundles at a Hamming distance of $1$ from bundle $b$ (see Section~$\ref{subsec:model-recommending}$). It then requires comparing $n\cdot (n-1)$ bundles when the interest bundle has size $1$, $(^n_2)\cdot(n-2)$ additional bundle pairs when the interest bundle size is $2$, and so on. ($(^n_k)$ Denotes the binomial coefficient.) Thus---worst case---there are $\sum^n_{k=1} (^n_k)\cdot (n-k) < (n \cdot 2^n)$ bundle comparisons necessary, which is significantly less than ${(2^n)}^2$. 

\subsection{Online Learning Mechanism}

In this paper we consider only the situation where the number of individual goods to be bundled is relatively small: i.e., $n\leq 10$. (With $n = 10$ there are $2^n -1=1023$ possible bundle configurations, so facilitating the search process among all those bundles is still highly valuable.) Since we only consider bundles within the neighborhood of an interest bundle, it is tractable, for relatively small values of $n$, to explicitly estimate the required conditional expectations online. Moreover, bargaining with one customer generally creates numerous training examples which can be used twice (i.e., $<b,b',p'-p>$ and $<b',b,p-p'>$ are both stored as separate training examples, see Section~$\ref{sec:bargaining_data}$). For small values of $n$ therefore, the learning mechanism can improve its estimation of the conditional expectations, even given relatively few customers who provide training examples.

Given the $k$ training examples $<<b,b',p_1-p'_1>,\ldots,<b,b',p_k-p'_k>>$, the online learning algorithm simply estimates $\Delta gt(b',b)$, the expected difference in gains from trade, resulting from changing from bargaining about bundle $b$ to bargaining about bundle $b'$ (given that a customer expressed an interest in bundle b), as the average of the training examples, i.e.,
\begin{equation}
\Delta \hat{gt}(b',b)=\frac{1}{k}\sum_{i=1}^{k}(p_i-p'_i).
\end{equation}
The danger of using $\Delta\hat{gt}(b',b)$ directly to sort the bundles in the neighborhood of the interest bundle $b$, is that the diversity of the trading example remains limited. Consequently, learning the correct ordering of the bundles is not possible. To allow for sufficient exploration the shop chooses with a probability $p(b,b',M^0)$ (with $M^0= Ng(b)$) a bundle $b'\in M^0$ to be first in the ordering of $Ng(b)$; once the first bundle in the ordering is determined, say $b^*$, with a probability $p(b,b',M^1)$ (with $M^1= Ng(b) \setminus\{b^*\}$) the shop chooses a bundle $b\in M^1$ to be second in the ordering, and so on. The probability $p(b,b',M)$ (with $M\subseteq Ng(b)$) is computed according to the softmax action selection rule (cf. \cite{Sutton/Barto:1998}), i.e.,
\begin{equation}
p(b,b',M) = \frac{e^{\lambda \cdot \Delta \hat{gt}(b',b)}}{\sum_{b''\in M} e^{\lambda \cdot \Delta \hat{gt}(b'',b)}}.
\end{equation}
where $\lambda$ determines the exploratory behavior of the mechanism. The greater $\lambda$, the less exploration will take place, i.e., the higher the probability that the bundle with the highest expected gains from trade will be picked first, the second highest second, and so on. Initially $\lambda$ is very small; it increases over time.

\section{Numerical Experiments}\label{sec:simulation}
In order to test the performance of our proposed mechanism, we implemented it computationally, and tested it against many simulated customers. Valuations for the shop and the customers were drawn from random distributions. First we describe customer preferences and how we implemented negotiations in the simulation, and then we present our experimental design and simulation results.

\subsection{Customer preferences}\label{subsec:customerPreferences}

The goods may be complementary, in which case the valuation of a bundle is higher than the sum of the valuations of the individual goods in the bundle. We model the possibility of complementarities by representing $v_c(b)$, the customer's valuation for a bundle $b$, as a (cubic) polynomial. If we let $N$ denote the collection of $n$ individual goods and the vector $x=(x(1), \ldots, x(n))$ the binary representation of a bundle $b$ (i.e., $x(i) =1$ if and only if $i\in b$), then
\begin{eqnarray}\label{eq:valuation}
v_c(b) & = & a_0 + \sum_{i \in N}a_i\cdot x(i) + \sum_{i,j\in N}a_{ij}\cdot x(i) \cdot x(j)+\nonumber\\
       &   & \sum_{i,j,k\in N}a_{ijk}\cdot x(i) \cdot x(j) \cdot x(j),
\end{eqnarray}
where $a_0$, $a_i$, $a_{ij}$, and $a_{ijk}$ (for $i,j,k\in N$) denote the constant, linear, quadratic, and cubic coefficients of the polynomial, respectively. The quadratic and cubic coefficients determine the extent to which complementarities exist between two and three goods. (Customers buy at most one instance of an individual good, hence we can ignore the possibility of complementarities between identical goods: i.e., $a_{ii}=a_{iii}=0$ for all $i\in N$.)

An individual customer's values for the various coefficients are randomly distributed. If $\vec{a}$ denotes an arbitrary instance of all these coefficients, then $\vec{a}$ gives rise to a multivariate normal distribution, i.e., $pr(\vec{a})\sim N[\vec{\mu},\mathbf{\Sigma }]$, where the vector $\vec{\mu} $ and the matrix $\mathbf{\Sigma}=[\sigma_{ij}]$ denote the means and (co)variances of the distribution. 

From Equation~$\ref{eq:valuation}$ (and the fact that $x(i)\in \{0,1\}$) it follows that we can obtain all bundle valuations $(v_c(b_1), \ldots, v_c(b_{2^n-1}))$ by applying a linear transformation on $\vec{a}$. Consequently, the corresponding probabilities $pr(v_c(b_1), \ldots, v_c(b_{2^n-1}))$ also form a multivariate normal distribution \cite{green:1993}. That is, we have
\begin{equation}
pr(v_c(b_1), \ldots, v_c(b_{2^n-1})) \sim N[\mathbf{T}\vec{\mu},\mathbf{T\Sigma T'}],
\end{equation}
where the matrix $\mathbf{T}$ specifies the linear transformation (the $j^{th}$ element in the $i^{th}$ row of $\mathbf{T}$ specifies whether or not the corresponding $i^{th}$ coefficient in $\vec{a}$ should contribute to the valuation of the $i^{th}$ bundle). 

\subsection{Modeling Negotiations}\label{subsec:simulation-modelingNegotiations}

\subsubsection{Time-dependent Strategy}

\newcommand{\tft}{\textsc{tft}}
\newcommand{\tdf}{\textsc{tdf}}
\newcommand{\tftm}{\textsc{tftm}}

For the customer (shop), the time-dependent bidding strategy is monotonically increasing (decreasing) in both the number of bidding rounds ($t$) and her (his) valuation. In particular, a bidding strategy is characterized by the gap the customer leaves between her initial offer and her valuation, and by the speed with which she closes this gap. The gap is specified as a fraction of the bundle valuation and it decreases over time as $\mathit{gap(t)} = \mathit{gap_{init}} \cdot \exp (-\delta t)$, so over time, she approaches the valuation of the bundle she is currently negotiating about. Note that changes in the gap are time-dependent, but not bundle-dependent! This strategy is therefore called ``time-dependent-fraction'' (\tdf). Almost the same holds for the shop's bidding strategy, \textit{mutatis mutandis}. The initial gap, $\mathit{gap_{init}}$, is set at $0.5$ for the customer and at $1.5$ for the shop, and we fix $\delta = 0.03$ for the shop as well as the customer, in order to reduce the number of jointly fluctuating parameters somewhat. Summing up, the customer (shop) starts her (his) bidding for a bundle at (one and a) half her (his) valuation, and her (his) bids gradually approach her (his) valuation.

\subsubsection{Tit-for-Tat Strategy}

The time-dependent strategy described above generates bids irrespective of what the opponent does. As an example of a strategy that responds to the opponent, we implemented a variant of tit-for-tat (\tft) \cite{axelrod:1984}. The initial `move' is already specified by $\mathit{gap_{init}}$ like in the \tdf-strategy. If in subsequent moves the utility level of the opponent's offer improves, then the same amount is conceded by the bargainer. Note that it is the increment in the utility level perceived by the bargainer (not the opponent). Furthermore, this perceived utility improvement can also be negative. To make the bidding behavior less chaotic, no negative concessions are made. That is, we used a so-called monotone version called tit-for-tat-monotone-fraction (\tftm) which can never generate a bid with a worse utility than the previous bid.

\subsection{Experimental Setup}\label{subsec:setup}

\begin{table*}[htb]
\caption{Comparison of the different methods $\mu$, $S$ and the benchmark $B$. Figures are averages across $10$ runs with different random seeds, and $12000$ customers per run. Standard deviations are given between brackets.}
\centering\begin{tabular}{|r|cc|cc|cc|}
\cline{2-7}
\multicolumn{1}{c|}{} & \multicolumn{6}{c|}{Methods}\\
\hline
\multicolumn{1}{|c|}{Performance} & \multicolumn{2}{c|}{$\mu$} & \multicolumn{2}{c|}{$S$} & \multicolumn{2}{c|}{$B$}\\
\multicolumn{1}{|c|}{Indicator} & \tdf & \tftm & \tdf & \tftm & \tdf & \tftm\\
\hline\hline
max.\ gains & \multicolumn{6}{c|}{$1202.81 (5.34)$}\\
min.\ gains & \multicolumn{6}{c|}{$-1023.61 (54.77)$}\\
gains $b_{\mathrm{init}}$ & \multicolumn{6}{c|}{$438.70 (15.45)$}\\
\hline
gains $b_{\mathrm{int}}$ & $763.55 (15.75)$ & $549.27 (4.23)$ & $826.18 (16.30)$ & $573.14 (5.81)$ & $697.74 (14.19)$ & $527.55 (11.61)$\\
gains $b_{\mathrm{final}}$ & $863.33 (4.57)$ & $797.45 (4.40)$ & $939.72 (8.60)$ & $875.98 (7.57)$ & $777.82 (8.59)$ & $727.81 (5.96)$\\
percentage & $0.85 (0.00)$ & $0.82 (0.00)$ & $0.88 (0.00)$ & $0.85 (0.00)$ & $0.81 (0.00)$ & $0.79 (0.00)$\\
rel. percentage & $0.52 (0.01)$ & $0.43 (0.01)$ & $0.61 (0.01)$ & $0.54 (0.01)$ & $0.41 (0.01)$ & $0.34 (0.01)$\\
rounds & $8.24 (0.60)$ & $5.16 (0.16)$ & $8.01 (0.65)$ & $4.71 (0.15)$ & $13.71 (0.93)$ & $7.37 (0.29)$\\
deals & $10314.20 (143)$ & $11024.50 (39)$ & $10340.60 (156)$ & $11114.20 (40)$ & $9171.80 (181)$ & $10496.50 (71)$\\
\hline
\end{tabular}
\label{table:results}
\end{table*}

\begin{figure*}[htb]
\resizebox{\textwidth}{!}{\includegraphics*[angle=270]{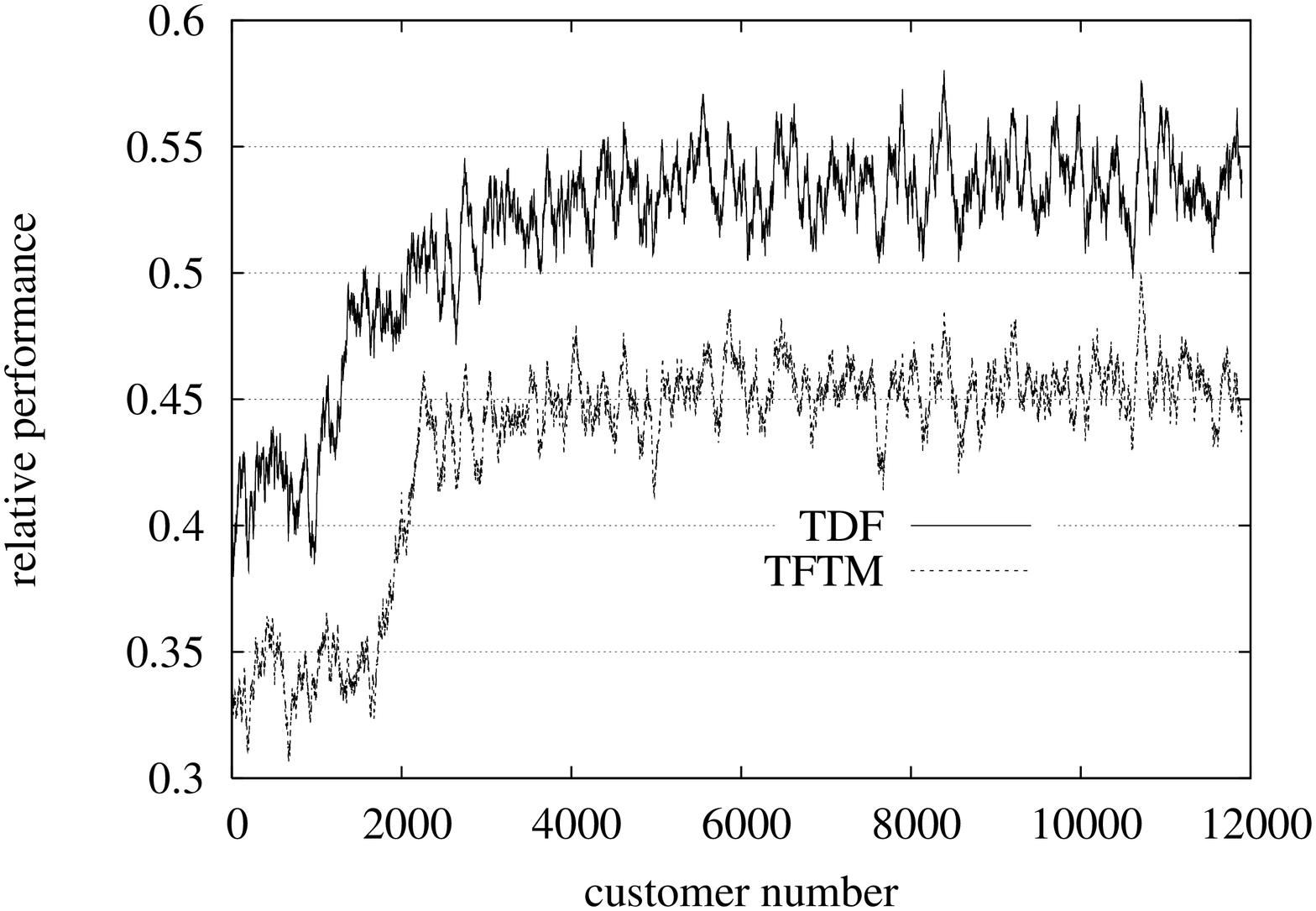} \includegraphics*[angle=270]{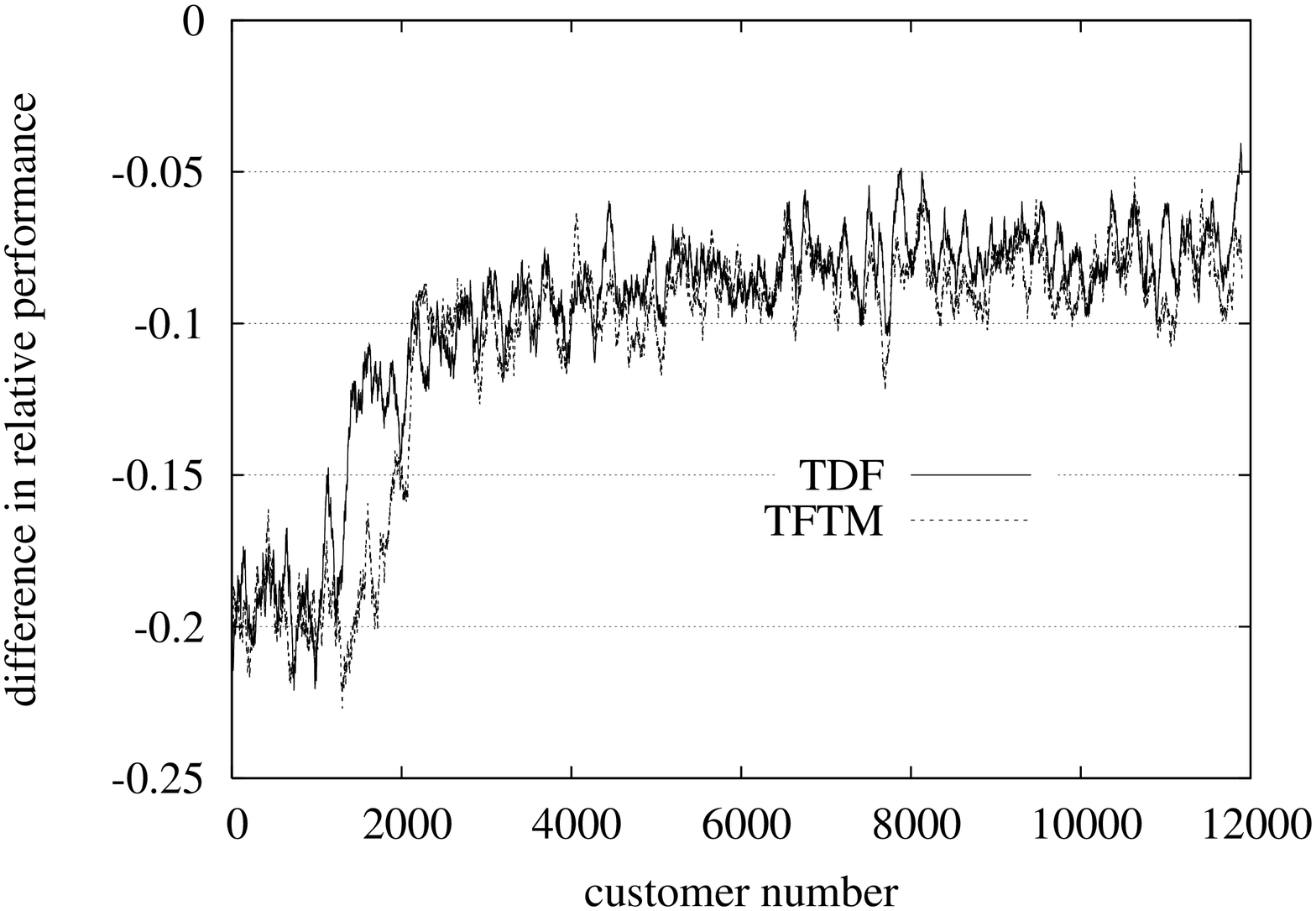}}
\caption{Relative performance of the $\mu$-system (on the left), measured by calculating the difference in gains from trade between the bundles $b_\mathrm{final}$ and $b_\mathrm{init}$, as a percentage of the difference between the maximum and the initial gains from trade. The shop uses the TDF strategy with $\delta = 0.03$, and the customers use either the TDF strategy (with $\delta = 0.03$), or the TFTM strategy (with $\delta=1$), as described in section~$\ref{subsec:simulation-modelingNegotiations}$. The graph on the right gives the difference in performance between $\mu$ and $S$. (Both graphs actually show the $100$-step moving averages.)}
\label{fig:relPercent}
\end{figure*}

\begin{figure*}[htb]
\resizebox{\textwidth}{!}{\includegraphics*[angle=270]{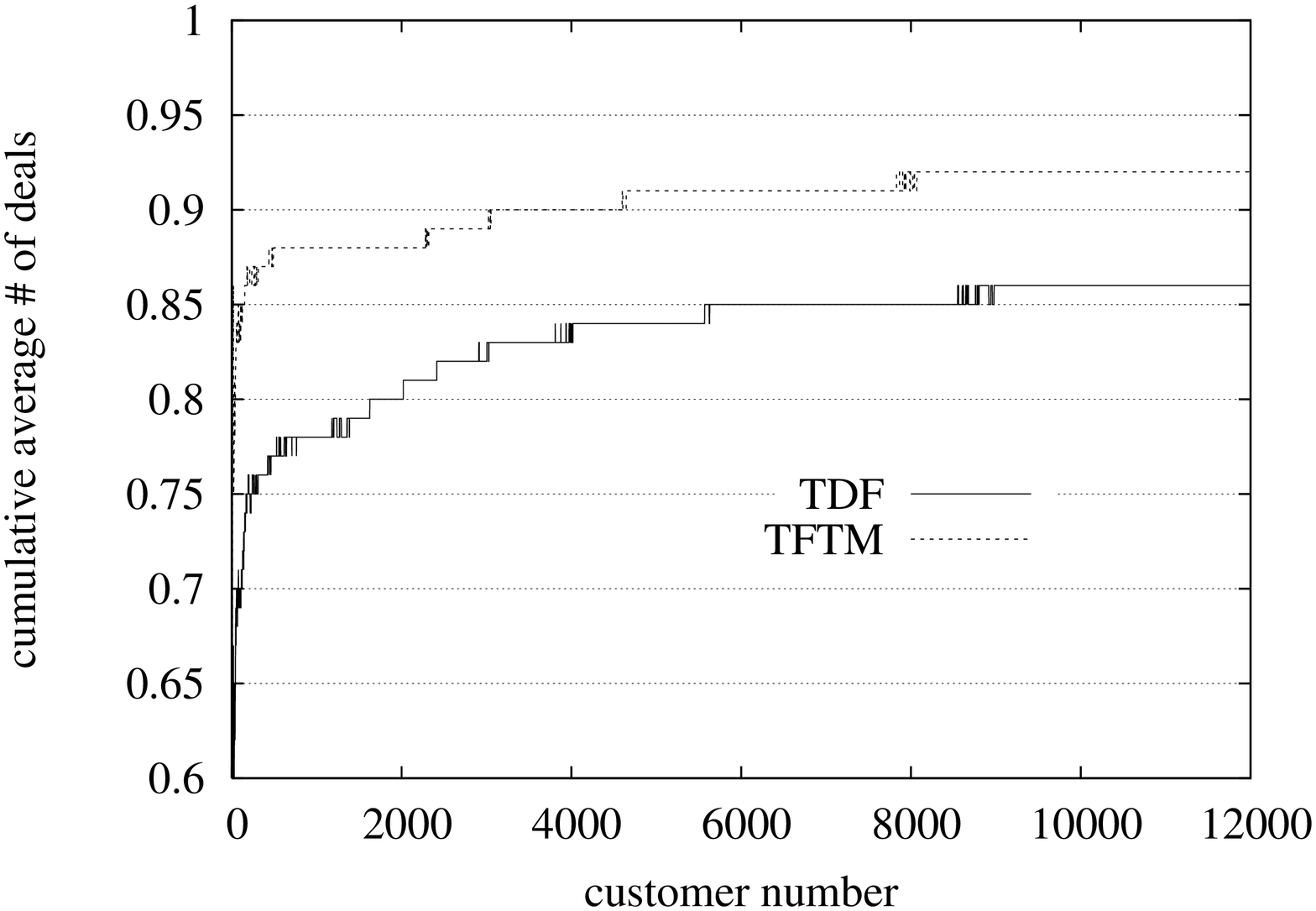} \includegraphics*[angle=270]{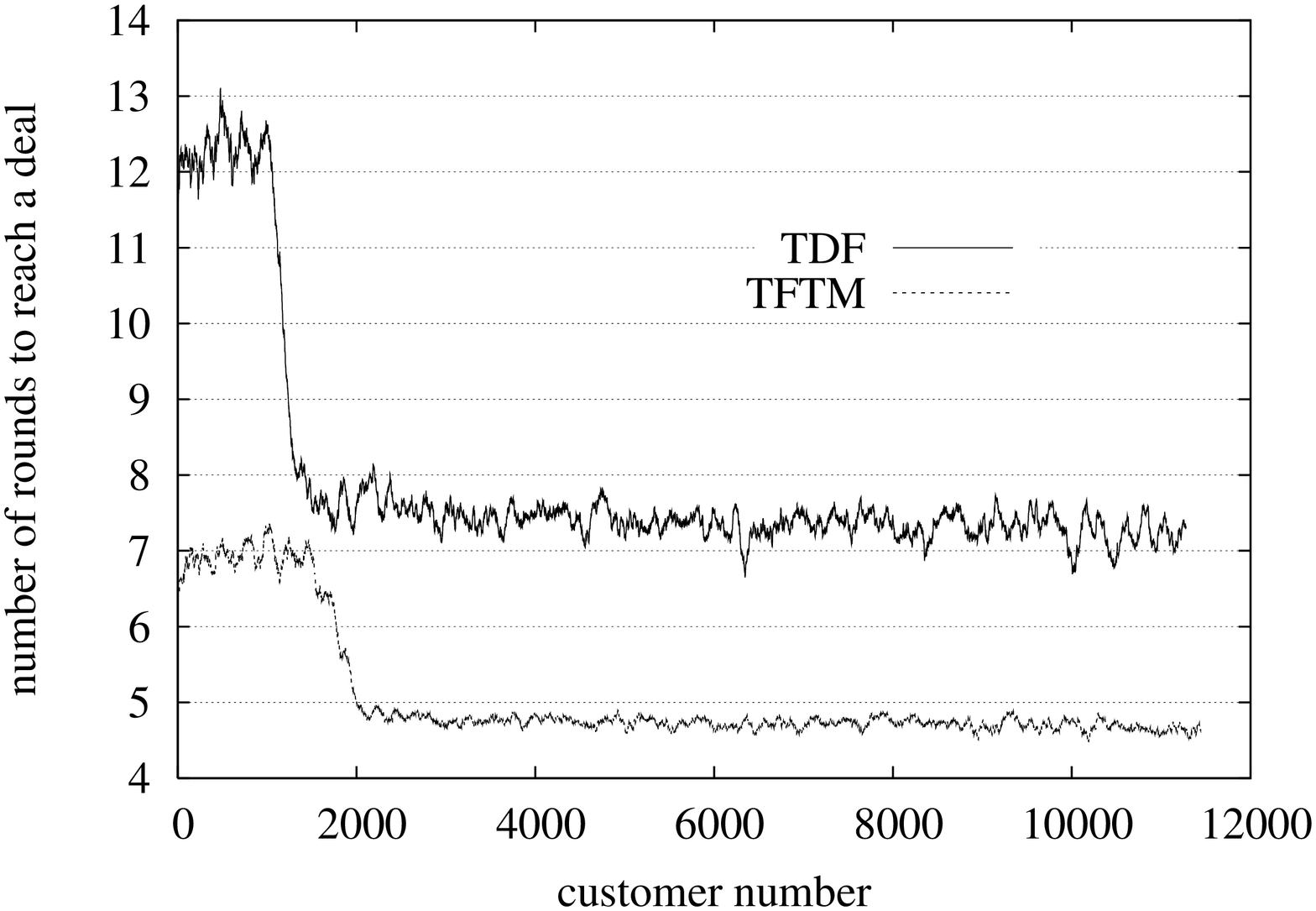}}
\caption{The cumulative average number of deals per customer, as attained by the $\mu$-system (on the left) and the $100$-step moving average of the number of rounds required to reach those deals (on the right). As in Figure~$\ref{fig:relPercent}$, the shop uses the TDF strategy and the customers use either the TDF or the TFTM strategy. The shop manages to reach deals with $11,381$ of the TDF-customers, and with $11,554$ of the TFTM-customers.}
\label{fig:dealsRounds}
\end{figure*}

In the computer experiments reported in this paper, we compare our new approach of having \emph{no a priori information} and learning customer preferences online (as discussed in Section~$\ref{sec:bargaining_data}$), to the one where---for example because of expert knowledge---the shop already knows the underlying joint probability distribution of all bundle valuations (see Section~$\ref{subsec:customerPreferences}$). The latter approach is also the one discussed and experimented with in more detail in \cite{Somefun/Etal:2004a}. In this approach, the shop directly derives the value of $E[v_c(b')|b]$ for each bundle pair; based on these values the shop computes the expected gains from trade and orders the bundles in the neighborhood of $b$ accordingly.

Besides comparing our new online procedure (referred to as $\mu$) with the previous method (called $S$), we also assess the relative performance of the system by performing the same series of experiments with a benchmark procedure (called $B$), which simply recommends a random bundle from the current bundle's neighborhood. That is, the benchmark does not base the order in which it recommends the next bundle on the estimated expected gains from trade like our system does.

In the experiments, the shop's bundle valuations are determined by applying a nonlinear bundle reduction. This means that the bundle price is generally less than the sum of the individual goods comprising the bundle. In order to prevent the trivial problem of customers wanting to buy all goods, the bundle reduction becomes $0$ for bundles which contain more than $3$ goods.

There are $10$ individual goods. We randomly generate the underlying probability density function $pr(\vec{a})$. In order to ensure sufficient difference in valuations between customers, however, we fix the correlation matrix (but not the covariance matrix). We randomly initialize the covariance matrix such that we can partition $N$ in $3$ subsets ($2$ of size $3$ and $1$ of size $4$). Selling a customer one of these $3$ subsets will often generate the highest gain from trade. (Roughly between $20$ and $40\%$ of the time this is the case, in the other cases $1$ or $2$ and sometimes more goods need to be added or removed.) To test the robustness of our procedure to quantitative changes in the underlying distributions we conducted a series of experiments with $10$ different distributions. For each of these settings we simulated negotiations between the shop, with randomly drawn valuations, which were kept constant across negotiations with $12,000$ customers, each with her valuations drawn randomly from the particular distribution used. To further test the robustness of our system, each customer was simulated using $2$ different negotiation strategies, as described in Section~$\ref{subsec:simulation-modelingNegotiations}$. (The shop always uses the \tdf\ strategy.)

To allow initiation of the negotiation process by the customer, we assume that the customer starts negotiating about an initial bundle $b_{\mathit{init}}$.  In order to give the shop some room for improvement, we initialize the customer's initial bundle by randomly selecting a bundle $b$ which, in binary representation, has a Hamming distance of $3$ to the bundle $b^*$ that is associated with the highest gains from trade.

\subsection{Results}\label{subsec:results}

The overall results of our experiments are listed in Table~$\ref{table:results}$. The numbers are averages over $12,000$ customers drawn from each distribution of valuations, and over $10$ different randomly generated distributions; standard deviations (across averages from the $10$ distributions) are listed between brackets. The maximum and minimum attainable gains from trade are determined by the current random distribution of valuations; they do not depend on the chosen strategy and method. Likewise, the bundle a customer wants to start negotiating about does not depend on the chosen strategy and method. Therefore, the average of these figures represented in the first $3$ rows are identical across all experiments---for each shop-customer interaction these figures are known even before the negotiation commences.

The remainder of the results is measured at the end of each shop-customer interaction, and subsequently averaged over all $10\cdot 12,000$ customers. The row for `gains $b_{\mathrm{int(erest)}}$' shows the gains from trade associated with the bundle of which the shop, at the end of the negotiation with each customer, thinks the customer is most interested in. This estimation is most accurately performed by the $S$-system, which is not surprising since it has direct access to the distribution underlying the customer's preferences---even though it does not know each individual customer's actual preferences. But the $\mu$-system, that has to do without this \emph{a priori} knowledge altogether, and instead has to learn about its customers' preferences online, does not do much worse, especially when compared to the benchmark system, $B$, in the rightmost columns.

The row labeled `gains $b_{\mathrm{final}}$' gives the gains from trade associated with the bundle that the shop and the customer were actually negotiating about at the end of the simulation, irrespective of whether that end was caused by the $98\%$ exogenous break-up probability, or by the fact that a deal was reached in the negotiation. The rows for `percentage' and `rel(ative) percentage' present the same results in a different way: `percentage' shows the shop's performance relative to the maximum attainable:
\[
\mbox{percentage}=\frac{(\mbox{gains }b_{\mathrm{final}}-\mbox{min.\ gains})}{(\mbox{max.\ gains}-\mbox{min.\ gains})},
\]
whereas `relative percentage' takes into account the starting bundle $b_{\mathrm{init}}$:
\[
\mbox{relative percentage}=\frac{(\mbox{gains }b_{\mathrm{final}}-\mbox{gains }b_{\mathrm{init}})}{(\mbox{max.\ gains}-\mbox{gains }b_{\mathrm{init}})}.
\]
Again, in both these rows, as in all the rows more generally, the $S$-system outperforms the $\mu$-system, which beats the $B$-system, but bear in mind the challenge for the $\mu$-system, as compared to the $S$-system, in terms of (dealing with the lack of) available aggregate knowledge. The rows labeled `rounds' and `deals' give the average number of rounds it took to reach a deal (whenever a deal was reached) and the average number of deals reached. An observation that can be made is that the shop seems to do better (in terms of gains from trade) when the customers use the \tdf-strategy than when they use \tftm, although in the former case the shop requires a higher number of rounds to reach deals, and reaches less deals, as compared to the latter case.

Figures~$\ref{fig:relPercent}$ and $\ref{fig:dealsRounds}$ illustrate the shop's learning process when using the $\mu$-system. The graph on the left in Figure~$\ref{fig:relPercent}$ shows (the $100$-step moving average of) the `relative percentage' from Table~$\ref{table:results}$, measured at the end of the negotiation with each of the $12,000$ customers, and averaged over the $10$ different preference-distributions. The increase over time, of the shop's aggregate knowledge of his customer's preferences is clearly visible, for both strategies used by the customers. 

The graph on the right in Figure~$\ref{fig:relPercent}$ shows the difference of these results between the $\mu$- and the $S$-systems, respectively; the $S$-system does better, but the $\mu$-system closes the gap as it learns more about its customers. Significantly, the difference between the plots for \tdf\ and \tftm\ disappears, indicating the robustness of the $\mu$-system to changes in the customers' negotiation strategies. So the $\mu$-system is clearly able to learn customers' preferences online, irrespective of the negotiation strategy used by those customers. However, the overall performance of the shop using the $\mu$-system together with his own negotiation strategy, \emph{is} dependent upon the customer's negotiation strategy. More specifically there is a trade-off in performance. Compared to \tftm, interacting with customers using \tdf\ results in higher gains from trade, less deals (see also Figure~$\ref{fig:dealsRounds}$), and more rounds to reach those deals. The explanation for these differences is that with \tftm\ customers will give in quicker; whenever the shop suggests a good alternative the amount the customer concedes equals the gains from trade plus the amount conceded by the shop (perceived by the customer). Consequently deals are reached more quickly. This also results in more deals being reached but goes at the expense of the gains from trade because the search process is shorter. 

\section{Conclusions and Future Work}\label{sec:discussion}
In this paper, we consider a form of multi-issue negotiation where a shop negotiates both the contents and the price of bundles of goods with his customers. To facilitate the negotiations of a shop, we develop a procedure that uses aggregate (anonymous) knowledge about \emph{many} customers in bilateral negotiations of bundle-price combinations with \emph{individual} customers. By online interpreting customers' responses to the shop's proposals for negotiating about alternative bundles, the procedure acquires the desired aggregate knowledge online; it requires no a priori information while respecting customers' privacy.

We conduct computer experiments with simulated customers that have \emph{nonlinear} preferences. We compare our new approach of having \emph{no a priori information} and learning about customer preferences online, to the one where---for example because of expert knowledge---the shop already knows the underlying joint probability distribution of all bundle valuations. The latter approach is also the one discussed and experimented with in more detail in \cite{Somefun/Etal:2004a}. Our experiments show how, over time, the performance of our procedure approaches that of our previous procedure, which already possesses the necessary aggregate knowledge. Both procedures significantly increase the speed with which deals are reached, as well as the number and the Pareto efficiency of the deals reached, as compared to a benchmark. Moreover, the experiments show that the new procedure is able to learn the necessary information online, irrespective of the negotiation strategy used by the customers.

\bibliographystyle{abbrv}
\bibliography{names,ec}

\begin{thebibliography}{10}

\bibitem{altinkemerJaisingh:2002-ieeewecwis}
K.~Altinkemer and J.~Jaisingh.
\newblock Pricing bundled information goods.
\newblock In {\em Proceedings of {IEEE-WECWIS'02}}, 2002.

\bibitem{axelrod:1984}
R.~Axelrod.
\newblock {\em The Evolution of Cooperation}.
\newblock Basic Books, New York, 1984.

\bibitem{Bakos/Brynjolfsson:1999}
Y.~Bakos and E.~Brynjolfsson.
\newblock Bundling information goods: Pricing, profits and efficiency.
\newblock {\em Management Science}, 45(12), December 1999.

\bibitem{Baumol/Etal:1987}
W.~Baumol, R.~Willig, and J.~Panzar.
\newblock {\em Contestable Markets and the Theory of Industry Structure}.
\newblock Dryden Press, 1987.

\bibitem{chuangSirbu:1999-iep}
J.~C.-I. Chuang and M.~A. Sirbu.
\newblock Optimal bundling strategy for digital information goods.
\newblock {\em Information Economics and Policy}, 11(2):147--176, 1999.

\bibitem{costerGustavssonOlssonRudstrom:2002-rpec}
R.~C{\"o}ster, A.~Gustavsson, T.~Olsson, and {\AA}.~Rudstr{\"o}m.
\newblock Enhancing web-based configuration with recommendations and
  cluster-based help.
\newblock In {\em Proceedings Workshop on Recommendation and Personalization in
  eCommerce at AH2002}, 2002.

\bibitem{Ehtamo2001}
H.~Ehtamo and R.~H{\"a}m{\"a}l{\"a}inen.
\newblock Interactive multiple-criteria methods for reaching pareto optimal
  agreements in negotiation.
\newblock {\em Group Decision and Negotiation}, 10:475--491, 2001.

\bibitem{faratinSierraJennings:2003}
P.~Faratin, C.~Sierra, and N.~R. Jennings.
\newblock Using similarity criteria to make issue trade-offs.
\newblock {\em Journal of Artificial Intelligence}, 142(2):205--237, 2003.

\bibitem{green:1993}
W.~H. Green.
\newblock {\em Econometric Analysis}.
\newblock Prentice Hall, New Jersey, 1993.

\bibitem{kephartFay:2000-acmec}
J.~O. Kephart and S.~A. Fay.
\newblock Competitive bundling of categorized information goods.
\newblock In {\em Proceedings of {ACM EC'00}}, 2000.

\bibitem{Klein2003}
M.~Klein, P.~Faratin, H.~Sayama, and Y.~Bar-Yam.
\newblock Negotiating complex contracts.
\newblock {\em Group Decision and Negotiation}, 12:111--125, 2003.

\bibitem{Mas-Collel/Etal:1995}
A.~{Mas-Collel}, M.~D. Whinston, and J.~R. Green.
\newblock {\em Mircoeconomic Theory}.
\newblock Oxford University Press, 1995.

\bibitem{molina:2001-wi}
M.~Molina.
\newblock An intelligent sales assistant for configurable products.
\newblock In {Zhong, N. (et al.)}, editor, {\em Proceedings Web Intelligence
  2001}, volume 2198 of {\em {LNAI}}, pages 596--600. Springer-Verlag, Berlin,
  2001.

\bibitem{resnickVarian:1997-cacm}
P.~Resnick and H.~R. Varian.
\newblock Recommender systems.
\newblock {\em Communications of the {ACM}}, 40(3):56--58, 1997.

\bibitem{Rubinstein:1982}
A.~Rubinstein.
\newblock Perfect equilibrium in a bargaining model.
\newblock {\em Econometrica}, 50(1):97--109, January 1982.

\bibitem{Schmalensee:1984}
R.~L. Schmalensee.
\newblock Gaussian demand and commodity bundling.
\newblock {\em Journal of Business}, 57(1):S211--S230, January 1984.

\bibitem{somefunGerdingBohtePoutre:2003-amec}
D.~J.~A. Somefun, E.~Gerding, S.~Bohte, and J.~A. La~Poutr{\'e}.
\newblock Automated negotiation and bundling of information goods.
\newblock In {J. A. Rodriguez-Aguilar (et al.)}, editor, {\em Proceedings of
  Agent Mediated Electronic Commerce V}, volume 3048 of {\em LNAI}.
  Springer-Verlag, Berlin, 2004.

\bibitem{Somefun/Etal:2004a}
D.~J.~A. Somefun, T.~B. Klos, and J.~A. La~Poutr{\'e}.
\newblock Negotiating over bundles and prices using aggregate knowledge.
\newblock In {\em {Proceedings $5^{th}$ International Conference on Electronic
  Commerce and Web Technologies}}, volume {(forthcoming)} of {\em LNCS}.
  Springer-Verlag, Berlin, 2004.

\bibitem{Somefun/Poutre:2003}
D.~J.~A. Somefun and J.~A. La~Poutr{\'e}.
\newblock Bundling and pricing for information brokerage: Customer satisfaction
  as a means to profit optimization.
\newblock In {\em Proceedings of Web Intelligence 2003}, pages 182--189. IEEE
  Computer Society, 2003.

\bibitem{sooLiang:2001-cia}
V.-W. Soo and S.-H. Liang.
\newblock Recommending a trip plan by negotiating with a software travel agent.
\newblock In M.~Klusch and F.~Zambonelli, editors, {\em Proceedings CIA 2001},
  volume 2182 of {\em LNAI}, pages 32--37. Springer-Verlag, Berlin, 2001.

\bibitem{Sutton/Barto:1998}
R.~S. Sutton and A.~G. Barto.
\newblock {\em Reinforcement Learning: An Introduction}.
\newblock Adaptive Computation and Machine Learning. The MIT Press, Cambridge
  Massachusetts, 1998.

\end{thebibliography}
\end{document}